\documentclass[conference]{IEEEtran}
\IEEEoverridecommandlockouts
\usepackage{lineno}
\usepackage{cite}
\usepackage{amsmath,amssymb,amsfonts}
\usepackage{amsmath}
\usepackage{amsthm}
\usepackage{graphicx}
\usepackage{textcomp}
\usepackage{xcolor}
\usepackage{graphicx}
\usepackage{float}
\usepackage{subfigure}
\usepackage{amsmath}
\usepackage{amsfonts,amssymb}
\usepackage{mathrsfs}
\usepackage{mathtools}
\usepackage{bm}
\usepackage{multirow}
\usepackage{array}
\usepackage{amssymb}
\usepackage{amsmath}
\usepackage{cite}
\usepackage{url}
\usepackage{xcolor}
\usepackage{cite,graphicx,amsmath,amssymb}
\usepackage{subfigure}
\usepackage{fancyhdr}
\usepackage{mdwmath}
\usepackage{mdwtab}
\usepackage{caption}
\usepackage{amsthm}
\usepackage{setspace}
\usepackage{bm}
\usepackage{algorithm}
\usepackage{mathtools}
\usepackage{dsfont}
\usepackage{bbm}
\usepackage{multicol}
\usepackage{algpseudocode}
\usepackage{framed}
\usepackage{tikz}
\usetikzlibrary{shapes.geometric, arrows, positioning}
\usepackage[nolist]{acronym}

\DeclareMathOperator*{\argmax}{argmax}

\newcommand{\mI}{\mathbf{I}}

\newcommand{\bv}{\mathbf{v}}
\newcommand{\br}{\mathbf{r}}
\newcommand{\bu}{\mathbf{u}}

\newcommand{\bx}{\mathbf{x}}

\newcommand{\bz}{\mathbf{z}}

\newcommand{\bl}{\mathbf{l}}

\newcommand{\setL}{\mathbb{L}}

\newcommand{\R}{\mathcal{R}}
\newcommand{\I}{\mathcal{I}}

\newcommand{\maD}{\mathcal{D}}

\newcommand{\tx}{\mathrm{tx}}
\newcommand{\rx}{\mathrm{rx}}
\newcommand{\ur}{\mathrm{u}}
\newcommand{\tg}{\mathrm{t}}
\newcommand{\ap}{\mathrm{ap}}
\newcommand{\subto}{\text{\normalfont s.t.} \;\; }

\newcommand{\dif}{\mathrm{d}}

\newcommand{\trp}{\mathsf{T}}
\newcommand{\her}{\mathsf{H}}
\newcommand{\set}[1]{\left\lbrace #1 \right\rbrace}

\newcommand{\brc}[1]{\left( #1 \right)}
\newcommand{\dbc}[1]{\left[ #1 \right]}
\newcommand{\norm}[1]{\left\Vert #1 \right\Vert}
\newcommand{\abs}[1]{\left\vert #1 \right\vert}

\newcommand{\Ex}[1]{\mathbb{E} \left\lbrace #1 \right\rbrace}

\def\BibTeX{{\rm B\kern-.05em{\sc i\kern-.025em b}\kern-.08em
    T\kern-.1667em\lower.7ex\hbox{E}\kern-.125emX}}
    \expandafter\def\expandafter\normalsize\expandafter{%
    \normalsize%
    \setlength\abovedisplayskip{4pt}%
    \setlength\belowdisplayskip{4pt}%
    \setlength\abovedisplayshortskip{2pt}%
    \setlength\belowdisplayshortskip{2pt}%
}
\begin{document}
\begin{acronym}
\acro{ap}[AP]{access point}
\acro{mimo}[MIMO]{multiple-input multiple-output}
\acro{los}[LoS]{line-of-sight}
\acro{isac}[ISAC]{integrated sensing and communications}
\acro{sic}[SIC]{successive interference cancellation}
\acro{snr}[SNR]{signal to noise ratio}
\acro{sinr}[SINR]{signal to interference and noise ratio}
\acro{bcd}[BCD]{block coordinate descent}
\acro{rzf}[RZF]{regularized zero-forcing}
\acro{awgn}[AWGN]{additive white Gaussian noise}
\acro{pas}[PASS]{\underline{p}inching \underline{a}ntenna \underline{s}y\underline{s}tem}
\acro{smi}[SMI]{sensing mutual information}
\acro{rcs}[RCS]{radar cross-section}
\end{acronym}

\title{
Low-Complexity Tuning of Pinching-Antenna Systems for Integrated Sensing and Communication
}
\author{
\IEEEauthorblockN{Saba Asaad}
\IEEEauthorblockA{University of Toronto\\
saba.asaad@utoronto.ca}
\and
\IEEEauthorblockN{Chongjun Ouyang}
\IEEEauthorblockA{Queen Mary University of London\\
c.ouyang@qmul.ac.uk}
\and
\IEEEauthorblockN{Zhiguo Ding}
\IEEEauthorblockA{Nanyang Technological University\\
	zhiguo.ding@ntu.edu.sg}
\and
\IEEEauthorblockN{Ali Bereyhi}
\IEEEauthorblockA{University of Toronto\\
ali.bereyhi@utoronto.ca}
\thanks{This work was supported in part by German Research Foundation (DFG).}
}

\maketitle

\begin{abstract}
Pinching antenna systems (PASSs) can dynamically adapt their transmit and receive arrays for sensing and communication in wireless systems. This work explores the potential of PASSs for integrated sensing and communication (ISAC) by proposing a novel PASS-aided ISAC design, in which pinching locations are adaptively adjusted to enable simultaneous sensing and data transmission with minimal interference. The proposed design introduces a \textit{bi-partitioning strategy} that allocates sensing power and tunes pinching locations with remarkably low computational complexity, allowing dynamic PASS tuning at high update rates. Numerical results demonstrate that the proposed approach achieves a significantly larger sensing-communication rate region compared to baseline designs at no noticeable cost. 
\end{abstract}
\section{Introduction}
\label{sec:intro}
The recently proposed \textit{\ac{pas}} has emerged as a promising solution to overcome the limitations of traditional antenna arrays in dynamic wireless environments \cite{suzuki2022pinching, ding2025flexible,LiuPASS}. By allowing electromagnetic waves to propagate within a low-loss dielectric waveguide and radiate through tunable pinching elements, \ac{pas} introduces a flexible analog front-end, which can adaptively reconfigure the radiation pattern \cite{bereyhi2025icc,bereyhi2025mimopass,ouyang2025array}. This reconfigurability enables the system to effectively suppress large-scale fading and establish strong \ac{los} links, making \ac{pas} particularly attractive for medium size indoor communication scenarios \cite{Fan2025PASS}.

Given its flexibility, \ac{pas} naturally aligns with the objectives considered in \ac{isac} systems \cite{Liu2022ISAC}. In \ac{isac} systems, transmit and receive arrays simultaneously exchange data and sensing signals, which often interfere. The ability of \ac{pas} to dynamically adjust the locations of its pinching elements offers a significant degree of freedom to spatially decouple sensing and communication channels, and thereby efficiently mitigating their mutual interference \cite{ISAC_PASS}. Several recent studies have analyzed this potential, reporting performance bounds that highlight substantial gains of \ac{pas}-aided \ac{isac} architectures. The studies in \cite{ISAC_PASS} and \cite{ISAC_PASS2} characterize the sensing-communication rate region with \ac{pas} transceiver, reporting a drastic expansion of the rate region under optimal \ac{pas} update. The work in   \cite{ISAC_PASS3} deploys \acp{pas} to boost the illumination power of the sensing signals in \ac{isac} systems, while guaranteeing a threshold communication Quality-of-Service (QoS). Although these lines of work elaborate the potential of \ac{pas}-aided transmission for \ac{isac}, their reported result remains idealistic, as they consider optimal tuning of the pinching locations on the transmit and receive waveguides. In practice, the realization of such potentials requires dynamic reconfiguration of the \ac{pas} structure at rates comparable to the channel coherence \cite{asaad2025dynamicEE}. This presents a challenge: designing a fast tuning algorithm that adjusts pinching locations rapidly while maintaining the enhanced performance. This work addresses this challenge by developing low-complexity algorithms for \ac{pas}-aided \ac{isac}.

\paragraph*{Contributions}
In this paper, we develop a low-complexity \textit{bi-partitioning} method for fast dynamic reconfiguration of \acp{pas}, when deployed for joint sensing and communication in \ac{isac} systems. The proposed method uses the geometry of the coverage area to partition the pinching elements into \textit{communication-centric} and \textit{sensing-centric} elements, and finds the optimal group-sizes by solving a reduced scalar optimization problem. 
The key contributions in this work are three-fold:
($i$) We formulate the \ac{pas}-aided \ac{isac} design task in both uplink and downlink cases as a multi-objective optimization problem, which dynamically adjusts the transmit and receive pinching locations for optimal trade-off between the spectral efficiency and sensing rate. ($ii$) We develop low-complexity algorithms for joint power allocation and location tuning in both downlink and uplink scenarios. The proposed algorithms optimize the pinching locations in a single step, i.e., with fixed complexity $\mathcal{O}(1)$, using a geometric approach that optimally splits the pinching elements into communication- and sensing-centric partitions. The former partition is mainly responsible for reconfiguring the communication channel while the latter adjusts the sensing environment. We refer to this technique as \textit{bi-partitioning} method. ($iii$) We validate our design through numerical experiments, comparing the proposed \ac{pas}-aided \ac{isac} design against the fixed-antenna baseline. Our investigation depicts that the proposed scheme can significantly enlarge the sensing-communication rate region as compared with baseline, while maintaining negligible computational cost. The results suggest that \ac{pas}-aided front-end can significantly enhance performance of multi-functional wireless systems. 

\paragraph*{Notation}
We show the scalars, vectors, and matrices with non-bold, bold lowercase, and bold uppercase letters, respectively. Expectation is denoted by $\Ex{\cdot}$, and the integer sets $\set{1, \ldots,n}$ and $\set{n, \ldots,m}$ are shortened as $[n]$ and $[n:m]$, respectively.

\section{System Model and Problem Formulation}
\label{sec:formulation}


Consider two dielectric waveguides extended over the $x$-axis at altitude $a$. Each waveguide is equipped with multiple pinching elements that can freely move across the $x$-axis. The waveguides are distanced $d$ on the $y$-axis with the first one being located at $y=0$ and the other at $y=d$. 
For simplicity, we refer to the first waveguide as the \textit{transmit \ac{pas}} and the other as \textit{receive \ac{pas}} in the sequel. The transmit and receive \acp{pas} are equipped with $N$ and $M$ pinching elements, respectively. The coordinate of the pinching element $n$ in the transmit \ac{pas} is given by $\bv^\tx \brc{\ell_n^\tx} = [\ell_n^\tx, 0, a]$, and the coordinate of the element $m$ on the receive \ac{pas} is $\bv^\rx \brc{\ell_m^\rx}= [\ell_m^\rx, d, a]$, where $0 \leq \ell_n^\tx \leq L^\tx$ and $0 \leq \ell_m^\rx \leq L^\rx$ with $L^\tx$ and $L^\rx$ being the length of the transmit and receive waveguides, respectively. 

The \ac{ap} aims to deploy the transmit-receive \ac{pas} pair to transmit communication signals to single-antenna users located at $\bu^\ur = [x_{\ur}, y_{\ur}, 0]$ and sense a target located at $\bu^\tg = [x_{\tg}, y_{\tg}, z_{\tg}]$. The user and target are located, such that a \ac{los} is available between them and the \acp{pas}.



\subsection{Downlink PASS-ISAC}
We first consider downlink \ac{pas}, in which the \ac{ap} transmits information signal to the user and senses the target from its reflection. Let $T$ denote the length of the time frame. At time $t\in [T]$, the \ac{ap} sends the signal $x[t]$ to the user. We consider symbol-level average transmit power constraint, i.e.,
\begin{align}
    \frac{1}{T} \sum_{t=1}^T  \Ex{\abs{x[t]}^2} \leq P^{\max},
\end{align}
for some maximum transmit power $P^{\max}$. Assuming ergodicity, we can write the constraint as
$\Ex{\abs{x[t]}^2} \leq P^{\max}$.

The signal received by the user at time $t$ is then given by
\begin{align}
    y_\ur [t] &= \sum_n g_\tx \brc{\ell_n^\tx, \bu^\ur} x[t]  + \varepsilon[t],
\end{align}
where $\varepsilon [t]$ is \ac{awgn} with zero mean and variance $\sigma_\ur^2$, and $g_\tx \brc{\ell, \bu}$ is the effective channel from the transmit \ac{pas} through an element at location $\ell$ to a receiver at coordinate $\bu$ and is given by
\begin{align}
    g_\tx \brc{\ell, \bu}
    =
    \frac{\beta \exp\set{ -j \kappa \brc{\maD_\tx \brc{\ell, \bu} + i_{\rm ref} \ell} }}{\sqrt{N} \maD_\tx \brc{\ell, \bu}}.
\end{align}
Here, $\beta$ captures the effective surface of the pinching element and shadowing, $i_{\rm ref}$ is the reflective index of the waveguide, $\kappa = 2\pi / \lambda$ is the wavenumber at wavelength $\lambda$, and $\maD_\tx \brc{\ell, \bu}$ denotes the distance between location $\ell$ on the transmit waveguide and coordinate $\bu$, i.e., $\maD_\tx \brc{\ell, \bu} = \norm{ \bv^\tx \brc{\ell} - \bu }$. For the sake of compactness, we define
\begin{align}
    G_\tx \brc{\bl^\tx, \bu}=\sum_n g_\tx \brc{\ell_n^\tx, \bu}.
\end{align}

In practice, user reflections are canceled, since users are registered units. The received signal contains unknown reflections from the target and can be written as
\begin{align}
    r [t] = \xi_\tg G (\bl^\tx, \bl^\rx, \bu^\tg)  
    x[t]  + \nu[t],\label{eq:r}
\end{align}
where $\nu [t]$ is zero-mean \ac{awgn} with variance $\sigma_{\rx, \rm d}^2$, and the scalar $\xi_\tg$ 
is the unknown \ac{rcs}. Following the \emph{Swerling-I} model, we assume the \ac{rcs} is relatively constant from pulse-to-pulse with an \emph{a prior} Rayleigh distributed amplitude. As a result,  $\xi_\tg$ is modeled as a complex zero-mean Gaussian random variable with variance $\rho^2_{\tg}$. In \eqref{eq:r}, the expresison $G (\bl^\tx, \bl^\rx, \bu)$ is further the transmit-to-receive \ac{pas} channel through a target at $\bu$ defined as 
\begin{subequations}
    \begin{align}
    G (\bl^\tx, \bl^\rx, \bu) &= \sum_m \sum_n g_\tx\brc{\ell_n^\tx, \bu} g_\rx \brc{\ell_m^\rx, \bu} \\
    &=G_\tx \brc{\bl^\tx, \bu}G_\rx \brc{\bl^\rx, \bu} \label{eq:GtxGrx},
\end{align}
\end{subequations}
with $g_\rx \brc{\ell, \bu}$ denoting the channel from an emitting target at point $\bu$ to the receive \ac{pas} through an element located at $\ell$ on the receive waveguide and is given by
\begin{align}
    g_\rx \brc{\ell, \bu}
    =
    \frac{\beta \exp\set{ -j \kappa \brc{\maD_\rx \brc{\ell, \bu} + i_{\rm ref} \ell} }}{\maD_\rx \brc{\ell, \bu}},
\end{align}
for $\maD_\rx \brc{\ell, \bu} = \norm{ \bv^\rx \brc{\ell} - \bu }$ and
\begin{align}
    G_\rx \brc{\bl^\rx, \bu}=\sum_m g_\rx \brc{\ell_m^\rx, \bu}.
\end{align}


\subsection{Uplink PASS-ISAC}
In uplink \ac{isac}, the user transmits its information signal $x[t]$ to the \ac{ap}, and the \ac{ap} broadcasts the sensing signal $z[t]$ via its transmit \ac{pas}, where the transmit power of the user and the \ac{ap} are limited by $P_{\rm c}$ and $P_{\rm s}$, respectively. 
The sensing signal is reflected by the target. Invoking the channel reciprocity and considering perfect cancellation of user reflections, the received signal by the \ac{ap} is given by
\begin{align}
    y_\ap [t] =\;  &G_\rx \brc{\bl^\rx, \bu^\ur} x[t] + \xi_\tg G (\bl^\tx, \bl_m^\rx, \bu^\tg)  z[t]  + \eta [t],
\end{align}
for $\eta [t] \sim \mathcal{CN} \brc{0,\sigma_{\rx, \rm u}^2}$. The \ac{ap} utilizes $y_\ap [t]$ to decode the transmitted information and estimate the target parameter $\xi_\tg$. 

\section{Communication and Sensing Metrics}
We now formulate the \ac{pas}-aided \ac{isac} design using the notions of 
\textit{spectral efficiency} 
and \textit{\ac{smi}} \cite{sensningMI, ISAC_MI}.


\subsection{Downlink ISAC}
In downlink \ac{isac}, the spectral efficiency is given by
\begin{align}
    \R_{\rm c}^{\rm d} \brc{\bl^\tx , P} = \log \brc{ 1 +
    \frac{
     P
    }{
     \sigma_\ur^2
    }\abs{G_\tx(\bl^\tx, \bu^\ur)}^2
    }
\end{align}
where $P = \Ex{\abs{x[t]}^2} \leq P^{\max}$. 

For sensing, the \ac{ap} estimates target parameter $\xi_\tg$ from its received reflections $r[1], \ldots, r[T]$. The \ac{smi} is defined as
\begin{align}
    \I_{\rm s} = \frac{1}{T} I\brc{ \br; \xi_\tg \vert \bu^\tg,  \bx },
\end{align}
where $\br = \dbc{r[1], \ldots, r[T]}^\trp$ collects reflected samples received at \ac{ap}, and $\bx = \dbc{x[1], \ldots, x[T]}^\trp$ is the downlink waveform. This metric measures the rate of sensing information collected from the environment, provided that knowledge on the transmit signal and target location is available, i.e., the target is tracked. 

To compute the \ac{smi}, we first write
\begin{align}
 I\brc{ \br; \xi_\tg \vert \bu^\tg , \bx } = h \brc{ \br \vert \bu^\tg , \bx } - h\brc{ \br\vert \xi_\tg,  \bu^\tg , \bx }.
\end{align}
Considering \eqref{eq:r}, we can use the Gaussianty of noise to write
\begin{align}
h\brc{ \br\vert \xi_\tg,\bu^\tg , \bx } = T\log 2 \pi e \sigma_{\rx,\rm d}^2 . \label{eq:h_0}
\end{align}
We next write 
\begin{align}
h \brc{ \br \vert \bu, \bx } = \int h \brc{ \br \vert \bx = \bx_0, \bu^\tg } p_\tx \brc{\bx_0} \dif\bx_0,\label{eq:h_1}
\end{align}
where $p_\tx \brc{\bx}$ is the distribution of $\bx$, and $\bu^\tg$ is known. For $\bx=\bx_0$, $\br$ is zero-mean complex Gaussian with covariance 
\begin{align}
    \boldsymbol{\Sigma} = \sigma_{\rx,\rm d}^2 \mI_T + 
    \rho_\tg^2 \abs{G (\bl^\tx, \bl^\rx, \bu)}^2  \bx \bx^\her.
\end{align}
With standard lines for derivation, we can conclude 
\begin{align}
\I_{\rm s} \hspace{-.5mm} = \hspace{-.5mm} \frac{1}{T}\mathbb{E}_{\bx} \set{ 
 \log \brc{1 \hspace{-.5mm} + \hspace{-.5mm} \frac{\rho_\tg^2}{\sigma_{\rx,\rm d}^2} 
   \abs{G (\bl^\tx, \bl^\rx, \bu^\tg)}^2  \norm{\bx}^2
   }
   }.
\end{align}

The exact expression for the \ac{smi} depends on  $p_\tx \brc{\bx}$. To derive a distribution-agnostic term, we use Jensen's inequality to evaluate a universal upper-bound as 
\begin{align}
\I_{\rm s} \hspace{-.5mm} \leq \hspace{-.5mm} \frac{1}{T} \set{ 
 \log \hspace{-.5mm} \brc{\hspace{-.5mm}1 \hspace{-.5mm} + \hspace{-.5mm} \frac{\rho_\tg^2}{\sigma_{\rx,\rm d}^2} 
   \abs{G (\bl^\tx, \bl^\rx, \bu^\tg)}^2  \Ex{\norm{\bx}^2}
   }
   \hspace{-1mm}}\hspace{-.5mm}.
\end{align}
Since
 $\Ex{\norm{\bx}^2}  = TP$,
we have 
 $\I_{\rm s} \leq \I_{\rm s}^{\rm d} \brc{\bl^\tx, \bl^\rx,P}$ with
\begin{align}
 \I_{\rm s}^{\rm d} &\brc{\bl^\tx, \bl^\rx,P} \nonumber\\
 &= \frac{1}{T} \set{ 
 \log \brc{1+ \frac{\rho_\tg^2TP}{\sigma_{\rx, \rm d}^2} 
   \abs{G (\bl^\tx, \bl^\rx, \bu^\tg)}^2  
   }
   }.
\end{align}
We consider the bound $\I_{\rm s}^{\rm d} \brc{\bl^\tx, \bl^\rx,P}$ as the sensing metric. 

\subsubsection*{Downlink Design Problem}
The design problem in this case is characterized by the following multi-objective optimization  
\begin{align}
    &\max_{\bl^\tx,\bl^\rx, P} 
    \set{
    \R_{\rm c}^{\rm d}  \brc{\bl^\tx , P}
    , \I_{\rm s}^{\rm d}  \brc{\bl^\tx, \bl^\rx,P}
    } \tag{$\mathcal{P}_{\rm d}$} \label{eq:Downlink}
    \\
    &\subto
    P \leq P^{\max},\bl^\tx \in \setL^\tx ,\bl^\rx\in \setL^\rx,
    \nonumber
\end{align}
where the feasible sets $\setL^\tx$ and $\setL^\rx$ are defined as
\begin{subequations}
    \begin{align}
    \setL^\tx &= \set{
    \ell_n^\tx: 0 \leq \ell_n^\tx \leq L^\tx \text{ and } \ell_n^\tx - \ell_{n-1}^\tx \geq \Delta
    },\\
    \setL^\rx &= \set{
    \ell_m^\rx: 0 \leq \ell_m^\rx \leq L^\rx \text{ and } \ell_m^\rx - \ell_{m-1}^\rx \geq \Delta
    },
\end{align}
\label{eq:LSet}
\end{subequations}
for some minimum spacing threshold $\Delta$.

\subsection{Uplink ISAC}
In uplink \ac{isac}, the sensing and communication signals interfere over-the-air. The \ac{ap} uses \ac{sic}: 
it first decodes the communication signal while treating the sensing signal as noise. The spectral efficiency is hence given by
\begin{align}
    &\R_{\rm c}^{\rm u}  \brc{\bl^\tx, \bl^\rx, P_{\rm c}, P_{\rm s}}
    \nonumber
    \\ &= 
    \log \brc{ 1 +
    \frac{
    \abs{G_\rx(\bl^\rx, \bu^\ur)}^2 P_{\rm c}
    }{
    \rho_\tg^2 \abs{ G(\bl^\tx, \bl^\rx, \bu^\tg)}^2 P_{\rm s} + \sigma_{\rx,\rm u}^2
    }
    },
\end{align}
where $P_{\rm c} = \Ex{\abs{x[t]}^2}$ and $P_{\rm s} = \Ex{\abs{z[t]}^2}$.

After decoding the information signal, the \ac{ap} cancels its impact by subtracting it from the received signal. Assuming negligible decoding error, the sufficient statistics computed after interference cancellation is given by
\begin{subequations}
    \begin{align}
    \hat{y}_\ap [t] &= y_\ap [t] - G_\rx \brc{\bl^\rx, \bu^\ur} s[t]  \\
    &= \xi_\tg G \brc{\bl^\tx, \bl^\rx, \bu^\tg} z[t]  + \nu [t].
\end{align}
\end{subequations}
Following the same lines of derivations as in downlink scenario, the \ac{smi} is derived as 
\begin{align}
 &\I_{\rm s} = \frac{1}{T} \mathbb{E}_{\bz}\set{
 \log \brc{1+ \frac{\rho_\tg^2}{\sigma_{\rx,\rm u}^2} 
   {\abs{G (\bl^\tx, \bl^\rx, \bu^\tg)}^2}  \norm{\bz}^2
   }},
\end{align}
where $\bz = \dbc{z\dbc{1}, \ldots, z\dbc{T}}$. Using Jensen's inequality, the \ac{smi} can be bounded from above by $\I_{\rm s}^{\rm u}\brc{\bl^\tx, \bl^\rx,P_{\rm s}}$, where
\begin{align}
 \I_{\rm s}^{\rm u}&\brc{\bl^\tx, \bl^\rx,P_{\rm s}} =\nonumber\\
 &\frac{1}{T}  \log \brc{1+ \frac{\rho_\tg^2}{\sigma_{\rx,\rm u}^2} 
   {\abs{G (\bl^\tx, \bl^\rx, \bu^\tg)}^2}  TP_{\rm s}
   }.
\end{align}

\subsubsection*{Uplink Design Problem}
The design problem in this case~is 
\begin{align}
    &\max_{\bl^\tx,\bl^\rx, P_{\rm s}, P_{\rm c}} 
    \set{
    \R_{\rm c}^{\rm u}  \brc{\bl^\tx, \bl^\rx, P_{\rm c}, P_{\rm s}}
    , \I_{\rm s}^{\rm u}  \brc{\bl^\tx, \bl^\rx, P_{\rm s}}
    }\tag{$\mathcal{P}_{\rm u}$} \label{eq:Uplink}
    \\
    &\subto
    P_{\rm s} \leq P_{\rm s}^{\max}, P_{\rm c} \leq P_{\rm c}^{\max},\bl^\tx \in \setL^\tx ,\bl^\rx\in \setL^\rx , \nonumber
\end{align}
for $\setL^\tx$ and $\setL^\rx$ defined in \eqref{eq:LSet}.

\section{ISAC System Design}
The optimal design is given by the Pareto front of the multi-objective optimization, whose characterization is not tractable. We develop an algorithm to approximate the Pareto front.
\subsection{PASS Design for Downlink ISAC}
Noting that both communication and sensing metrics are increasing in $P$, we can conclude that for any point in the Pareto front the optimal $P^\star = P^{\max}$. One can further use the monotonic behavior of the logarithm function and rewrite the downlink design problem in \eqref{eq:Downlink} as
\begin{align}
    &\max_{\bl^\tx,\bl^\rx} 
    \set{
    \abs{G_\tx(\bl^\tx, \bu^\ur)}^2
    , \abs{G (\bl^\tx, \bl^\rx, \bu^\tg)}^2  
    } \label{eq:Downlink1}
    \\
    &\subto
    \bl^\tx \in \setL^\tx ,\bl^\rx\in \setL^\rx.
    \nonumber
\end{align}
Noting that the reflection channel decomposes as
\begin{align}
    \abs{G (\bl^\tx, \bl^\rx, \bu^\tg)}^2 =\abs{G_\tx \brc{\bl^\tx, \bu^\tg} }^2 \abs{G_\rx \brc{\bl^\rx, \bu^\tg}}^2,
\end{align}
we can decouple \eqref{eq:Downlink1} into two marginal problems: a receiver design problem, which solves the single-objective optimization
\begin{align}
    &\max_{\bl^\rx} 
   \abs{G_\rx \brc{\bl^\rx, \bu^\tg}}^2 
    \subto
    \bl^\rx\in \setL^\rx,
    \label{eq:DownlinkRx}
\end{align}
and a transmit design problem which solves
\begin{align}
    &\max_{ \bl^\tx \in \setL^\tx } 
    \set{
    \abs{G_\tx(\bl^\tx, \bu^\ur)}^2
    , \abs{G_\tx \brc{\bl^\tx, \bu^\tg} }^2  
    }. \label{eq:DownlinkTx}
\end{align}

\subsubsection*{Receive PASS Design}
For \eqref{eq:DownlinkRx}, it is straightforward to find the solution geometrically. In fact, optimal $\bl^\rx$ should have the minimum distance to the target and result in in-phase superposition of all signals received by the pinching elements. Without spacing restriction, both these conditions are satisfied by setting all elements at $x_\tg$. Considering the minimum distance constraint, this can be efficiently approximated by distancing the elements by the minimum multiply of the waveguide that is larger than the minimum spacing threshold $\Delta$. Assuming that $\Delta \leq \lambda$, we can approximate accurately the solution of \eqref{eq:DownlinkRx} by setting $\bl^\rx = \bl^\rx_\star$, where
\begin{align}
    \ell_{m,\star}^\rx = x_\tg + \brc{m - \frac{M+1}{2}} \lambda.
    \label{eq:RX_Loc}
\end{align}
Here, we set a mass center at $x_\tg$ and place the~elements~symmetrically around it with every two neighbors distanced $\lambda$. 




\subsubsection*{Transmit PASS Design}
To solve the multi-objective problem in \eqref{eq:DownlinkTx}, we invoke the scalarization scheme, in which we optimize a weighted sum of the objectives. This means that we find a Pareto optimal solution by setting
\begin{align}
    &\max_{ \bl^\tx \in \setL^\tx } 
\omega_1
    \abs{G_\tx(\bl^\tx, \bu^\ur)}^2
+ \omega_2 \abs{G_\tx \brc{\bl^\tx, \bu^\tg} }^2 .
\label{eq:DownlinkTx2}
   \end{align}
Unlike the receive \ac{pas} design, the transmit pinching locations should maximize the sum channel gain to the user and to the target. The exact solution is not straightforward; however, it can be efficiently approximated by a simple geometric trick, which we refer to as \textit{bi-partitioning trick} in the sequel. 

\subsubsection*{Bi-Partitioning Trick}
We divide transmit pinching elements into two sets: a set with $N_1$ elements and the other with $N-N_1$ elements. We center the first part close to the user, i.e. uniformly around a mass center at $x_\ur$ with spacing $\lambda$, and the second part close to the target, i.e. uniformly around $x_\tg$ with spacing $\lambda$. Assuming that elements in each part are spaced for in-phase superimposition at both user and target, we have
\begin{align}
\abs{G_\tx(\bl^\tx, \bu^\ur)}^2
   &\approx 
   \beta N \abs{
   \frac{N_1}{N D_{\ur,1}} \zeta_{\ur}^1
   +\frac{N -N_1}{N D_{\ur,2}} \zeta_{\ur}^2
   }^2,
\end{align}
where $D_{\ur,1} = \maD_\tx \brc{x_\ur, \bu^\ur} \leq D_{\ur, 2} = \maD_\tx \brc{x_\tg, \bu^\ur}$ and $\zeta_{\ur}^1$ and $\zeta_{\ur}^2$ capture phase-shifts of first and second set, respectively. Defining $\alpha = N_1/N$ as the ratio of elements serving the user and $\zeta_{\ur} = \zeta_{\ur}^2/\zeta_{\ur}^1$ as the phase-shift difference, we can write
\begin{align}
\abs{G_\tx(\bl^\tx, \bu^\ur)}^2
   &\approx 
   \beta N \abs{
   \frac{\alpha}{D_{\ur,1}}
   +\frac{1-\alpha}{D_{\ur,2}} \zeta_{\ur}
   }^2.
\end{align}
Similarly, we can approximate $\abs{G_\tx(\bl^\tx, \bu^\tg)}^2$ as
\begin{align}
\abs{G_\tx(\bl^\tx, \bu^\tg)}^2
   &\approx 
   \beta N \abs{
   \frac{\alpha}{D_{\tg,2}}\zeta_{\tg}
   +\frac{1-\alpha}{D_{\tg,1}} 
   }^2,
\end{align}
where $D_{\tg,1} = \maD_\tx \brc{x_\tg, \bu^\tg} \leq D_{\tg, 2} = \maD_\tx \brc{x_\ur, \bu^\tg}$, and $\zeta_{\tg}$ captures the phase-shift difference.

We then approximate the optimal partitioning of the pinching elements by setting $N_1$ to $N_1^\star \approx \lfloor N \alpha^\star \rfloor$, where $\alpha^\star$ is the solution of 
   $\max_{0\leq \alpha \leq 1} F\brc{\alpha}$,
with $F\brc{\alpha}$ being 
\begin{align}
   F\brc{\alpha} = 
    \abs{
   \frac{\alpha}{D_{\ur,1}}
   +\frac{1-\alpha}{D_{\ur,2}}\zeta_{\ur}
   }^2 \hspace{-.7mm} + 
   \frac{\omega_2}{\omega_1}
    \abs{
   \frac{\alpha}{D_{\tg,2}}\zeta_{\tg}
   +\frac{1-\alpha}{D_{\tg,1}}
   }^2.
\end{align}
The locations on the transmit \ac{pas} are then set to
\begin{align}
    \ell_{n,\star}^\tx =
    \begin{cases}
        x_\ur + \brc{n - \frac{N_1^\star+1}{2}} \lambda &n \in [N_1^\star]\\
        x_\tg + \brc{n - \frac{N+N_1^\star+1}{2}} \lambda &n \in [N_1^\star+1: N]\\
    \end{cases}
    .\label{eq:TX_Loc}
\end{align}
The final algorithm is summarized in Algorithm~\ref{alg1}.

\begin{algorithm}[t]
\caption{Downlink ISAC-PASS Design}
\label{alg1}
\begin{algorithmic}[1]
\State Choose $(\omega_1,\omega_2)$ for a sensing-communication trade-off
\State Set pinching locations at the receive \ac{pas} to \eqref{eq:RX_Loc}
\State Set $\alpha^\star \gets \argmax_{0 \leq\alpha\leq 1} F\brc{\alpha}$ and $N_1 = \lfloor N \alpha^\star \rfloor$
\State Set pinching locations at the transmit \ac{pas} to \eqref{eq:TX_Loc}
\State \Return $\bl^\rx$, $\bl^\tx$, and $P = P^{\max}$
\end{algorithmic}
\end{algorithm}

\subsection{PASS Design for Uplink ISAC}
For the uplink case, we first note that $P_{\rm c}$ only shows up in the communication metric. Noting that this metric is increasing in $P_{\rm c}$, it is concluded that at any Pareto optimal point, we have $P_{\rm c} = P_{\rm c}^{\max}$. Considering this, the objectives in \eqref{eq:Uplink} are jointly optimized with respect to the sensing power and pinching locations. Using the scalarization with some weights $\omega_1$ and $\omega_2$, we can reformulate the uplink \ac{isac} problem as
\begin{align}
&\max_{P_{\rm s} \leq P_{\rm s}^{\max},\bl^\tx \in \setL^\tx ,\bl^\rx\in \setL^\rx}
     S_{\rm ul} \brc{\bl^\rx, \bl^\tx, P_{\rm s}}.
     \label{eq:Uplink2}
\end{align}
with $S_{\rm ul} \brc{\bl^\rx, \bl^\tx, P_{\rm s}}$ given in \eqref{eq:S_ul} at the top of the next page. 

\begin{figure*}[t!]
\begin{align}
S_{\rm ul} \brc{\bl^\rx, \bl^\tx, P_{\rm s}} = 
     \omega_1
    \log \brc{ 1 +
    \frac{
    \abs{G_\rx(\bl^\rx, \bu^\ur)}^2 P_{\rm c}^{\max}
    }{
    \rho_\tg^2 \abs{ G(\bl^\tx, \bl^\rx, \bu^\tg)}^2 P_{\rm s} + \sigma_{\rx,\rm u}^2
    }
    }
 + \frac{\omega_2}{T} 
 \log \brc{1+ \frac{\rho_\tg^2}{\sigma_{\rx,\rm u}^2} 
   {\abs{G (\bl^\tx, \bl^\rx, \bu^\tg)}^2}  TP_{\rm s}
   }\label{eq:S_ul}
\end{align}
\hrule
\end{figure*}

The problem in \eqref{eq:Uplink2} is NP-hard. 
We hence approximate its solution using the \ac{bcd} method, which alternates among: (i) joint sensing power optimization and transmit \ac{pas} tuning, i.e. optimizing $\bl^\tx$, and (ii) the receive \ac{pas} design, i.e. optimizing $\bl^\rx$. In the sequel, we present a tractable solution to each of these marginal problems.

\subsubsection*{Sensing Power and Transmit PASS Design}
The marginal problem in this case reduces to
\begin{align}
\max_{\bl^\tx, P_{\rm s}}
     S_{\rm ul} \brc{\bl^\rx, \bl^\tx, P_{\rm s}} \; \; 
     \subto
    P_{\rm s} \leq P_{\rm s}^{\max}, \bl^\tx \in \setL^\tx,
\end{align}
where $\bl^\rx$ is treated as fixed. Using \eqref{eq:GtxGrx}, one can decompose the effective channel gain $\abs{G(\bl^\tx,\bl^\rx, \bu^\tg)}$ as the product of $\abs{G_\tx(\bl^\tx, \bu^\tg)}$ and $\abs{G_\rx(\bl^\rx, \bu^\tg)}$, where the former term is only coupled by $P_{\rm s}$. We hence define the auxiliary variable $Q =\abs{G_\tx(\bl^\tx, \bu^\tg)}^2 P_{\rm s}$, and rewrite the marginal problem as
\begin{figure*}[t!]
\begin{align}
\tilde{S}_{\rm ul} \brc{\bl^\rx, Q} = 
     \omega_1
    \log \brc{ 1 +
    \frac{
    \abs{G_\rx(\bl^\rx, \bu^\ur)}^2 P_{\rm c}^{\max}
    }{
    \rho_\tg^2 \abs{ G_\rx(\bl^\rx, \bu^\tg)}^2 Q + \sigma_{\rx,\rm u}^2
    }
    }
 + \frac{\omega_2}{T} 
 \log \brc{1+ \frac{\rho_\tg^2}{\sigma_{\rx,\rm u}^2} 
   {\abs{G_\rx (\bl^\rx, \bu^\tg)}^2}  TQ
   }\label{eq:tilde_S_ul}
\end{align}
\hrule
\end{figure*}
\begin{align}
\max_{Q}
     \tilde{S}_{\rm ul} \brc{\bl^\rx, Q} \; \; 
     \subto
    Q \leq Q^{\max},
\end{align}
for $\tilde{S}_{\rm ul}$ given in \eqref{eq:tilde_S_ul} at the top of the next page, where $Q^{\max} = \abs{G_\tx(\bl^\tx_\star, \bu^\tg)}^2 P_{\rm s}^{\max}$ is the maximum scaled sensing powers achieved when setting $\bl^\tx = \bl^\tx_\star$ with
\begin{align}
    \ell_{n,\star}^\tx = x_\tg + \brc{n - \frac{N+1}{2}} \lambda,
    \label{eq:Up_TX_Loc}
\end{align}
for $n \in \dbc{N}$. It is straightforward to show that the solution is 
\begin{align}
    Q^\star =
    \begin{cases}
        Q^* &  Q^* \in \dbc{0, Q^{\max} }\\
        \displaystyle\argmax_{Q \in \set{0,Q^{\max}}} \tilde{S}_{\rm ul} \brc{\bl^\rx, Q}
        &  Q^* \notin \dbc{0, Q^{\max} }
    \end{cases},
    \label{eq:Qstar}
\end{align}
where 
$Q^* = \brc{q_0 - \sigma_\rx^2}/\brc{\rho_\tg^2 \abs{ G_\rx(\bl^\rx, \bu^\tg)}^2}$ 
for $q_0 \geq 0$ that is the non-negative solution to
\begin{align}
\omega_2 q^2 + \abs{G_\rx(\bl^\rx, \bu^\ur)}^2 P_{\rm c}^{\max} \brc{q - \varrho } = 0,
\end{align}
with $\varrho=\omega_1 \sigma_{\rx,\rm u}^2 \brc{1-T}$.

Considering the above solution, any choice of $\bl^\tx$ and $P_{\rm s}$ that satisfies $Q^\star =\abs{G_\tx(\bl^\tx, \bu^\tg)}^2 P_{\rm s}$ is a solution to the marginal problem. A straightforward choice is given by setting $\bl^\tx = \bl^\tx_{\star}$ and $P_{\rm s} = Q^\star /  \abs{G_\tx(\bl_\star^\tx, \bu^\tg)}^2$. 

\subsubsection*{Receive PASS Design}
Unlike the transmit \ac{pas}, the pinching locations at the receive \ac{pas} impact the sensing and communication metrics differently. For sensing enhancement, the elements should be located close to the target, while the spectral efficiency improves by locating the elements close to the user. We hence invoke the bi-partitioning idea once again: we set $M_1 = \lfloor \alpha M\rfloor$ elements to be located close to the user and the remaining to be located close to the target, i.e., we set 
\begin{align}
    \ell_{m}^\rx =
    \begin{cases}
        x_\ur + \brc{m - \frac{M_1+1}{2}} \lambda &m \in [M_1]\\
        x_\tg + \brc{m - \frac{M+M_1+1}{2}} \lambda &n \in [M_1+1: M]\\
    \end{cases}
    .\label{eq:RX_LocUp}
\end{align}
In this case, we can write $\abs{G_\rx\brc{\bl^\rx, \bu^\ur}}^2 \approx M^2 \hat{E}(\alpha, \bu^\ur) $ and $\abs{G_\rx\brc{\bl^\rx, \bu^\tg}}^2 \approx M^2 \hat{E}(\alpha, \bu^\tg) $, where
\begin{subequations}
\begin{align}
    \hat{E}(\alpha, \bu^\ur) &= \abs{\frac{\alpha}{D_{\ur,1}} +\frac{{1-\alpha}}{D_{\ur,2}} \zeta_{\ur} }^2,\\
    \hat{E}(\alpha, \bu^\tg) &= \abs{\frac{{1-\alpha}}{D_{\tg,1}}
    + \frac{\alpha}{D_{\tg,2}}\zeta_{\tg}
    }^2.
\end{align}
\end{subequations}
Here, we set $D_{\ur,1} = \maD_\rx \brc{x_\ur, \bu^\ur} \leq D_{\ur, 2} = \maD_\rx \brc{x_\tg, \bu^\ur}$ and $D_{\tg,1} = \maD_\rx \brc{x_\tg, \bu^\tg} \leq D_{\tg, 2} = \maD_\rx \brc{x_\ur, \bu^\tg}$, and define $\zeta_{\ur}$ and $\zeta_{\tg}$ to capture the phase-shift differences. The optimal ratio $\alpha^\star$ is approximated by solving 
    $\max_{ 0 \leq \alpha \in \leq 1 } \hat{S}_{\rm ul} \brc{\alpha}$, 
where 
\begin{align}
	\hat{S}_{\rm ul} \brc{\alpha} =\;& 
	\omega_1
	\log \brc{ 1 +
		\frac{
			{\hat{E}(\alpha, \bu^\ur)} P_{\rm c}^{\max} M^2
		}{
			\rho_\tg^2 Q^\star M^2 { \hat{E}(\alpha, \bu^\tg)}  + \sigma_{\rx,\rm u}^2
		}
	}\nonumber
	\\& + \frac{\omega_2}{T} 
	\log \brc{1+ \frac{\rho_\tg^2TQ^\star M^2}{\sigma_{\rx,\rm u}^2} 
		{{\hat{E}(\alpha, \bu^\tg)}}  
	}. \label{eq:hat_S_ul}
\end{align}
 The receive pinching locations are then set according to \eqref{eq:RX_LocUp} for $\alpha = \alpha^\star$. The final algorithm is summarized in Algorithm~\ref{alg2}.





\begin{algorithm}[t!]
\caption{Uplink ISAC-PASS Design}
\label{alg2}
\begin{algorithmic}[1]
\State Choose $(\omega_1,\omega_2)$ for a sensing-communication trade-off
\Repeat
\State Set pinching locations at the transmit \ac{pas} to \eqref{eq:Up_TX_Loc}
\State Compute $Q^\star$ via \eqref{eq:Qstar} and $\alpha^\star$. Set $M_1 = \lfloor M \alpha^\star \rfloor$
\State Set pinching locations at the receive \ac{pas} to \eqref{eq:RX_LocUp}
\Until{converges}
\State \Return $\bl^\rx$, $\bl^\tx$, $P_{\rm s}$, and $P_{\rm c} = P^{\max}$
\end{algorithmic}
\end{algorithm}

\subsection{Complexity and Convergence}
Both downlink and uplink algorithms impose minimal computational complexity on the system. In the downlink case, Algorithm~\ref{alg1} imposes $\mathcal{O}\brc{1}$ complexity, as it tunes the pinching locations via the bi-partitioning technique. Similarly, Algorithm~\ref{alg2} imposes computational complexity of order $\mathcal{O}\brc{T}$, where $T$ is the number of \ac{bcd} iterations. Through numerical experiments, we observe that the \ac{bcd} loop in Algorithm~\ref{alg2} rapidly converges, often after two or three \ac{bcd} iterations.

\section{Numerical Results}
We validate the proposed design through numerical experiments. 
We consider a setting,  
where both the user and sensing target are uniformly distributed within a rectangle centered at the origin, with side lengths $D_x$ and $D_y=8$ m along the $x$- and $y$-axes, respectively. The waveguides are deployed at a height of $a=3$ m, distanced with $d=4$ m. Both waveguides are aligned through the $x$-axis, i.e., $L^\tx = L^\rx = D_x$. 

Unless stated differently, the system parameters are set as follows: the carrier frequency is $f_{\rm{c}}=28$ GHz, both \acp{pas} are equipped with $N=M$ elements, effective refractive index is set to $n_{\rm{eff}}=1.4$, signal frame length is set to $T=5$, \ac{rcs} coefficient strength, i.e., its variance, is set to $\rho_\tg^2=10$, grid search is performed by resolution $10^4$, the minimum inter-element distance is set to $\Delta={\lambda}/{2}$, downlink power budget is set to $P^{\max}=10$ dBm, uplink communication and sensing power budgets are set to $P_{\rm{c}}^{\max}=P_{\rm{s}}^{\max}=5$ dBm, and noise variance is $\sigma_{\rm{c}}^2=\sigma_{\rm{s}}^2=-114$ dBm. In iterative algorithms, the pinching elements are initially positioned uniformly along the waveguides. For comparison, the proposed \ac{pas}-aided \ac{isac} configuration (PASS-ISAC) is evaluated against a baseline in which all antennas are fixed near the center of the waveguides and arranged linearly with half-wavelength spacing. Each antenna is connected to an analog phase shifter to implement conventional \emph{analog beamforming}, where the beamforming weights are optimized using the method proposed in \cite{wang2024beamfocusing}.

\begin{figure*}[!t]
\centering
    \subfigure[Downlink ISAC with $N=M=20$.]
    {
        \includegraphics[height=0.22\textwidth]{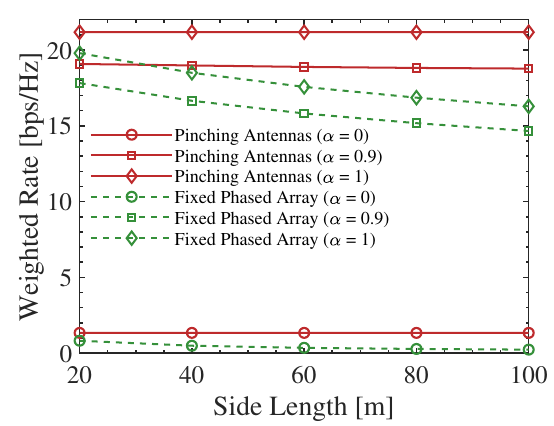}
	   \label{Figure_DL_Weighted_Rate_Side_Length}
    }
    \subfigure[Uplink ISAC with $N=M=10$.]
    {
        \includegraphics[height=0.22\textwidth]{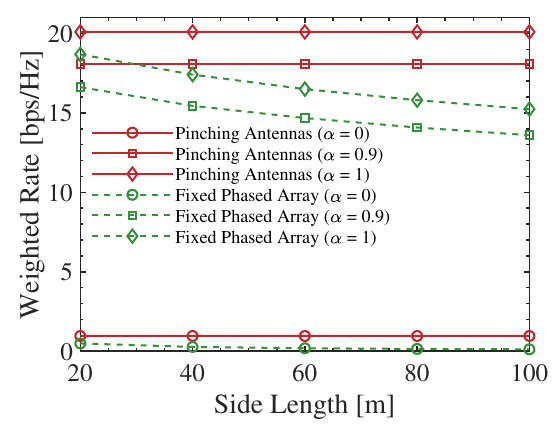}
	   \label{Figure_UL_Weighted_Rate_Side_Length}
    }
      \subfigure[Downlink ISAC at $D_x=50$.]
        {
    	\includegraphics[height=0.22\textwidth]{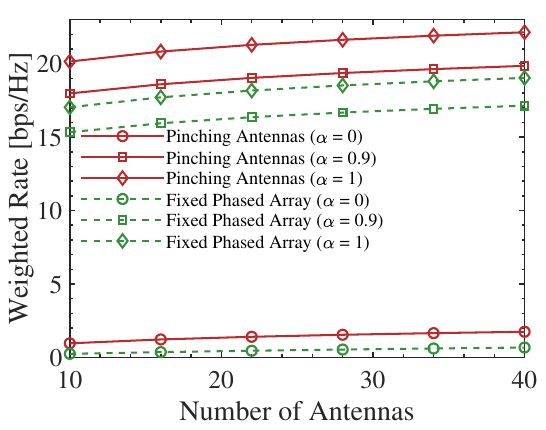}
    		   \label{Figure: Weighted_Rate_Antenna_Number}
    	    }
\caption{Weighted rate vs. the side length $D_x$ and the number of elements in downlink and uplink.}
\vspace*{-.2cm}
\label{Figure: Weighted_Rate_Side_Length}
\end{figure*}

As the first set of experiments, we investigate the impact of region size and \ac{pas} density on the weighted communication-sensing rate, i.e. the weighted sum of spectral efficiency and \ac{smi}. Here, we set the communication weight to $\alpha$ and the sensing rate to $1-\alpha$. Note that the weighting factor $\alpha$ determines the trade-off between communication and sensing. Specifically, $\alpha=0$ and $\alpha=1$ correspond to the \textit{sensing-centric} and \textit{communication-centric} design, respectively.

Figs. \ref{Figure_UL_Weighted_Rate_Side_Length} and \ref{Figure_DL_Weighted_Rate_Side_Length} plot the weighted rate against the side length $D_x$ for different weighting factors. As observed, for both downlink and uplink \ac{isac}, the proposed \ac{pas}-aided architecture consistently achieves a higher weighted rate as compared with conventional phased-array-based system. As the side length $D_x$ increases, the weighted rate achieved by the PASS-ISAC architecture remains nearly constant. This robustness arises due to the dynamic adjustability of the pinching elements along the dielectric waveguide, which minimizes the propagation distance between the radiating elements and the user and sensing target. In contrast, in the fixed phased-array architecture, all antennas are centrally located within the service region, and hence as the side length grows, the average distance between the array and both user and target increases, leading to more severe path loss and a substantial decline in the weighted rate. This explains why the performance advantage of PASS-ISAC over traditional architectures becomes more pronounced for larger service regions. It is worth noting that in the two extreme cases of $\alpha = 1$ and $\alpha=0$, the proposed PASS-ISAC architecture achieves higher communication and sensing rates than the conventional fixed phased-array system across all weighting configurations, demonstrating its superiority for joint communication and sensing in large-area deployments. Fig.~\ref{Figure: Weighted_Rate_Antenna_Number} depicts the weighted communication-sensing rate for downlink ISAC against pinching elements $N=M$. 
Across all configurations, the proposed PASS-ISAC consistently outperforms the conventional architecture. This follows from the ability of PASS to dynamically construct user and target adaptive strong \acp{los} by repositioning the elements. 

\begin{figure}[!t]
\centering
        \includegraphics[height=0.28\textwidth]{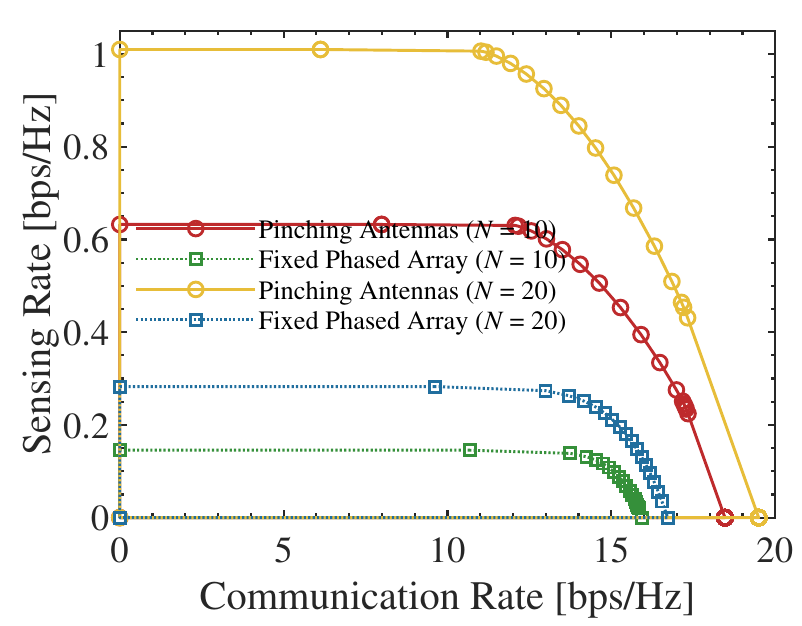}
\caption{Uplink ISAC rate region for side length $D_x=40$.}\vspace*{-.5cm}
\label{Figure: Rate_Region}
\end{figure}
Fig.~\ref{Figure: Rate_Region} presents the uplink communication-sensing rate region obtained by varying $\alpha$ from $0$ to $1$. As shown, the proposed PASS-ISAC achieves a significantly larger rate region compared to the baseline. Notably, the rate region of the conventional array is \textit{entirely contained} within that of PASS, demonstrating the ability of PASS-ISAC in achieving a more favorable communication-sensing trade-off. The results further indicate that the achievable rate region can be further expanded by increasing the number of elements.
\section{Conclusions}
This work proposed a low-complexity \ac{isac} design. The design uses \textit{bi-partitioning} to tune receive and transmit pinching locations in one shot, resulting in $\mathcal{O}(1)$ complexity for \ac{pas} tuning. Numerical experiments show that the proposed design can significantly expand the \ac{isac} rate region at no sensible computation cost, suggesting that \ac{pas}-aided \ac{isac} designs can be efficiently deployed in limited range wireless network with \ac{los} links, e.g., indoor systems. This highlights potentials of \acp{pas} for the next generation of wireless systems with multiple integrated functionalities. 

\bibliographystyle{IEEEtran} 
\bibliography{ref}
\end{document}